\documentclass{ws-p8-50x6-00}

\def\bd{\begin{document}} \def\ed{\end{document}}
\def\bmp{\begin{minipage}} \def\emp{\end{minipage}}
\def\bcc{\begin{center}} \def\ecc{\end{center}}     \def\npg{\newpage}
\def\beq{\begin{equation}} \def\eeq{\end{equation}} \def\hph{\hphantom}
\def\be{\begin{equation}} \def\ee{\end{equation}} \def\r#1{$^{[#1]}$}
\def\n{\noindent} \def\ni{\noindent} \def\pa{\parindent} 
\def\hs{\hskip} \def\vs{\vskip} \def\hf{\hfill} \def\ej{\vfill\eject} 
\def\cl{\centerline} \def\ob{\obeylines}  \def\ls{\leftskip}
\def\underbar#1{$\setbox0=\hbox{#1} \dp0=1.5pt \mathsurround=0pt
   \underline{\box0}$}   \def\ub{\underbar}    \def\ul{\underline} 
\def\f{\left} \def\g{\right} \def\e{{\rm e}} \def\o{\over} \def\d{{\rm d}} 
\def\vf{\varphi} \def\pl{\partial} \def\cov{{\rm cov}} \def\ch{{\rm ch}}
\def\la{\langle} \def\ra{\rangle} \def\EE{e$^+$e$^-$}
\def\bitz{\begin{itemize}} \def\eitz{\end{itemize}}
\def\btbl{\begin{tabular}} \def\etbl{\end{tabular}}
\def\btbb{\begin{tabbing}} \def\etbb{\end{tabbing}}
\def\beqar{\begin{eqnarray}} \def\eeqar{\end{eqnarray}}
\def\\{\hfill\break} \def\dit{\item{-}} \def\i{\item} 
\def\bbb{\begin{thebibliography}{9} \itemsep=-1mm}
\def\ebb{\end{thebibliography}} \def\bb{\bibitem}
\def\bpic{\begin{picture}(260,240)} \def\epic{\end{picture}}
\def\akgt{\noindent{Acknowledgements}}
\def\fgn{\noindent{\bf\large\bf Figure captions}}
\def\wp{W$^+$ \hs1pt } \def\wn{W$^-$ \hs1pt }
\def\epn{e$^+$e$^-$\hs1pt } \def\qqbar{q$\bar{\rm q}$} \def\ycut{y_{\rm cut}} 
\def\thcut{\theta_{\rm cut}}

\begin{document}

\title{A Monte Carlo Study on the Identification of Quark and Gluon Jets
\footnote{This work is supported in part by the NSFC under project 19975018}}

\author{Yu Meiling and Liu Lianshou}

\address{Institute of Particle Physics, Huazhong Normal University, 
Wuhan 430079\\
email: liuls@iopp.ccnu.edu.cn, yuml@iopp.ccnu.edu.cn}

\maketitle
\cl{\large Abstract}
\abstracts{
Three jets events in the \epn collisions at 91.2 GeV are investigated
using both HERWIG and JETSET Monte Carlo generators. The angles between 
the three jets are used to identify the quark and gluon jets.
The analysis at parton level is done to ensure the reasonableness
of this method and an angular cut is ultilized to improve the purity of this 
identification. 
The multiplicity inside the identified 
quark or gluon jets agree with the QCD predictions qualitatively.
}

In the theory of QCD, unobserved quarks and gluons are confined by color force.
They eventually hadronize into observed final state jets, which in a sense 
inherit the information about original partons.   In order to 
study the difference of nonpertubative hadronization processes between quarks 
and gluons, it is important to identify whether a final state jet is originated 
from a quark or gluon.  Many works have been done on this 
topic, see for example~\cite{MDPLB}~\cite{TASSO}~\cite{AMY}. 
Recently mostly used way are the 
so called b-tagged events in which b (and anti-b) quarks are well tagged 
and the rest jet is 
certainly a gluon jet~\cite{DELPHI}~\cite{OPAL}~\cite{ALEPH}. 
This method, apart from its high accuracy, is restricted in quark flavour
and is inapplicable to the gluon jets produced together with the 
light-quark jets.  

In this paper we analyse the events of \epn\ annihilations at 91.2 GeV 
generated by HERWIG~5.9~\cite{HERWIG} and JETSET~7.4~\cite{JETSET}
to find a method for the identification of quark and gluon jets
without flavour restriction.
Durham jet algorithm~\cite{DURHAM} is used to form 
jets both at parton level and at hadron level.
$R_2$ and $R_3$ are 
two and three jets fraction, respectively.
Distributions of $R_2$, $R_3$ at parton level and hadron level,
are obtained by varying $\ycut$ value from 
10$^{-4}$ to 10$^{-1}$, cf. Fig.1, in which
there exists a $\ycut$ value ($\log{(\ycut)}=-2.5$),  where  
$R_3$ both at parton level and at hadron level get to the maximum. 
We choose this $\ycut$ for all the later analysis.

We assume that the three jets in one event are coming from quark, antiquark 
and radiated gluon, respectively. Because of energy-momentum conservation, 
the three jets in one event must lie in a plan.  
These three jets are ordered according to the 
angles between two neighbouring jets. Jet-1 is defined as the jet opposite to 
the smallest angle $\theta_1$ and jet-3 is opposite to the largest angle
$\theta_3$. A three-jet event topology is depicted in Fig.2. 
 
\begin{picture}(230,230)
\put(-20,100)
{\epsfig{file=Fig/ycut11.epsi,width=140pt,height=120pt}
}
\put(150,120)
{\epsfig{file=Fig/topo.epsi,width=180pt,height=80pt}
}
\end{picture}
\vs-3.5cm
\n{\small {Fig. 1 \ \ two and three jets fraction as a
\hskip1.4cm Fig.2 \ \ Three-jet event topology

\n function of $\ycut$ at parton and hadron level. }
 }

In QCD,
the gluon radiated from one of the parent quark is expected to have 
relatively small energy and close to its parent quark, so in the 
above jet plan jet-1 might be a quark jet and jet-3 a gluon jet~\cite{AMY}. 
At parton level quark and gluon can be precisely identified. We select
three jets events by requiring two jets in one event contain 
opposite flavor q and $\rm \bar q$ seperately and the other one 
contains only gluon or \qqbar\  pairs. 

The distributions of ordered three angles at parton level
are shown in figure 3, in which
the smallest angle $\theta_1$  are well seperated from the largest one
$\theta_3$, which means that one of the quark jets are unambiguous  defined,
but the large overlap of $\theta_2$ and $\theta_3$ might cause 
wrong identification between the gluon jet and the other quark jet.
The efficiency of quark or gluon jet identification at parton level 
as a function of $\theta_3-\theta_2$ of each event is calculated by
comparing the type of jets determined from the angular rule, {\it i.e.}
assuming that gluon jet is always opposite to the largest angle,
to the original parton type. 
The results are in Fig.4, which shows that the larger the angular 
difference, the higher the efficiency.

In case of final state hadrons, events with 
{ $\vert \theta_1 + \theta_2 + \theta_3 - 360^\circ \vert < 1^\circ$}
are selected to satisfy the plannar request. This cut elimilates
about 26$\%$ events. 

In the application of the angular rule to final state hadrons, the 
direction of hadron jets might not be exactly the direction of the
original partons since the effect of hadronization will change 
the angles between jets. The events with three jets both at parton level
and at hadron level are selected to estimate the deviation. 
Of each event, hadron level ordered jet $i$
is matched to parton level ordered jet $i$ and the distribution of 
the angular differences 

\cl {parton $\theta_i$ - hadron $\theta_i \ \ \ \ ( i =  1, 2, 3)$} 

\n{are shown in Fig. 5.}

\vs0.5cm
\begin{picture}(200,230)
\put(-20,110)
{\epsfig{file=Fig/p3ang.epsi,width=160pt,height=120pt} }
\put(150,110)
{\epsfig{file=Fig/efficiency.epsi,width=160pt,height=120pt} }
\end{picture}
\vs-3.5cm
\n{\small Fig. 3 \ Distribution of angles between 
\hskip1cm Fig. 4 \ Efficiency of jet identification\\ 
jets at parton level 
\hs3.8cm as a function of the angular

\hs6cm difference between the largest angle 

\hs6cm and the medium one at parton level. }

\begin{center}
\begin{picture}(200,230)
\put(-70,90)
{\epsfig{file=Fig/phangdif1.epsi,width=100pt,height=120pt}
}
\put(50,90)
{\epsfig{file=Fig/phangdif2.epsi,width=100pt,height=120pt}
}
\put(160,90)
{\epsfig{file=Fig/phangdif3.epsi,width=100pt,height=120pt}
}
\end{picture}
\end{center}
\vs-3.1cm
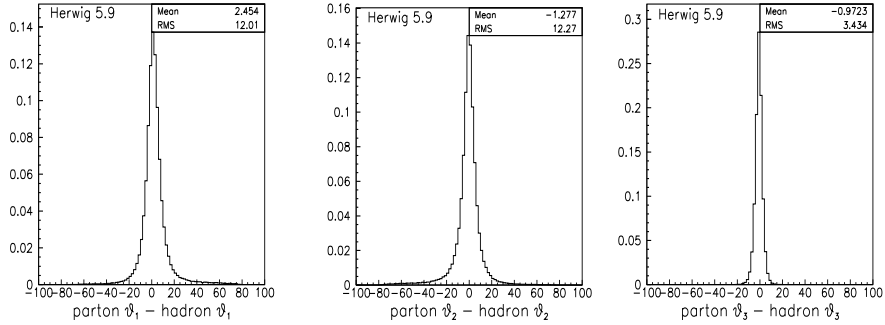
\cl{\small {Fig. 5 \ \ Corresponding angle difference between jets of parton
and jets of hadron. }
 }
\vs0.4cm
In these plots the largest angle $\theta_3$ which is 
supposed to opposite to gluon jet is the most stable one. It indicates
that the effects of hadronization on the relative direction of two quarks
are small, while the fluctuation on the direction of radiated gluon is 
much larger. On the average, 
the angle deviation between parton level and hadron level
is less than 3$^\circ$.

Based on the efficiency and angle deviation analyses given above, we 
require an angular cut $\thcut=20^\circ$ for $\theta_3 -\theta_2$ in 
hadron jet analysis and retain only the events with 
$\theta_3 -\theta_2 > \thcut$. This cut rejects about 50$\%$ events.    
Within the selected hadron events, we study the purity of angular rule.
Firstly, apply angular rule to hadrons to get quark and gluon jets sample.
Secondly, partons are forced into three jets and jets type is defined as the
efficiency study.
Thirdly, associate each hadron jet with a parton jet by finding the combination
which minimizes the sum of the angular difference between them and 
define pure events by requiring
hadron jet has the same jet type as its associated parton jet.
Purity which is defined as the number of pure events over the number of all 
selected hadron events is 63.8$\%$  for Herwig~5.9 and 83.2$\%$ for Jetset~7.4. 

Jets from two jets events are taken as quark jets which 
are compared with the three ordered jets from three jets events.
Because of the higher color charge of the gluon, it is expected
that gluon jets are of 
higher multiplicity than quark jets of the same energy~\cite{BG}.
In this paper, the multiplicity $n$ of the charged particles inside jet 
are investigated.
Jet energy is defined as the total energy of all the 
charged particles in the jet.  
The mean multiplicity $\la n \ra$ 
of jet as a function of jet energy are shown 
in Fig. 6 for HERWIG 5.9 and JETSET 7.4 respectively. 
Jet-1 and jet-2 which are identified as quark jet have the same behavior
as quark jets of two jets events, jet-3 which is expected to be a gluon jet
is well seperated from the other three samples and 
has higher mean multiplicity.

\begin{center}
\begin{picture}(200,230)
\put(-70,70)
{\epsfig{file=Fig/hmeann2.epsi,width=150pt,height=150pt}
}
\put(110,70)
{\epsfig{file=Fig/jhmeann2.epsi,width=150pt,height=150pt}
}
\end{picture}
\end{center}
\vs-2.6cm
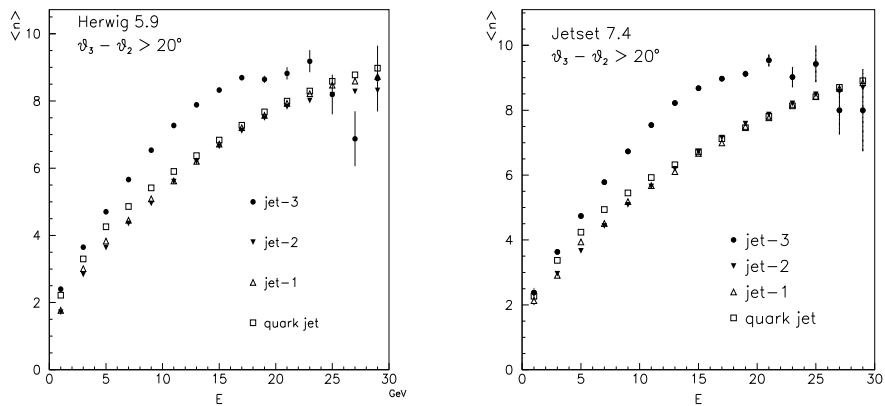
\n{\small {Fig. 6 \ \ Mean multiplicity and transverse momentum of jets
vs. jet energy, with angle cut 20 degree. }
 }
\vs0.4cm

In conclusion, based on the idea that the radiated gluon jet 
is close to its parent quark and from the idea of Parton-Hadron-Duality 
that
the direction of hadron jet is the same as parton jet, we use the angles
between final state hadron jets to identify quark or gluon jet. 
Two angular cuts are applied. The first one is planary cut, rejecting
the events with 
$\vert \theta_1 + \theta_2 + \theta_3 - 360^\circ \vert > 1^\circ$ 
The second one is the angular-difference cut, requiring
$\theta_3 -\theta_2 > \thcut$ with $\thcut = 20^\circ$. 
After these cuts the jet opposite to the largest angle $\theta_3$
is taken as the gluon jet.  The property of the gluon
and quark jet obtained in this way show obvious differences.
The higher mean multiplicity  of 
gluon jet sample agrees with  QCD prediction,
indicating that this angular method of jet identification
 is capable of distinguishing gluon 
jet from quark jet. In addition, there is no restriction on quark flavor
and no requirement on the jets in one event to have similar energy.
So it is relatively easy to carry out in experiment with high statistics.

\section*{Acknoledgements} 
 The authors thank Professor Sijbrand de Jong 
for helpful discussion.

\def\J#1#2#3#4{{#1} {\bf #2} (#3) #4}
\def\NCA{\em Nuovo Cimento} \def\NIM{\em Nucl. Instrum. Methods}
\def\NIMA{{\em Nucl. Instrum. Methods} A} \def\NPB{{\em Nucl. Phys.} B}
\def\PLB{{\em Phys. Lett.}  B} \def\PRL{\em Phys. Rev. Lett.}
\def\PRD{{\em Phys. Rev.} D} \def\ZPC{{\em Z. Phys.} C}
\def\PRE{{\em Phys. Rev.} E} \def\PRC{{\em Phys. Rev.} C}
\def\CPC{{\em Computer Physics Communications}}

\end{document}